\documentclass[%
 aip,
 amsmath,amssymb, 
 reprint,%
]{revtex4-1}

\usepackage{graphicx}
\usepackage{dcolumn}
\usepackage{bm}

\usepackage[utf8]{inputenc}
\usepackage[T1]{fontenc}
\usepackage{xcolor,color,soul}
\usepackage{mathptmx}
\usepackage{etoolbox}
\usepackage{upgreek} 
\usepackage{multirow}
\usepackage{algorithm2e}

{
\makeatletter
\def\frontmatter@thefootnote{%
 \altaffilletter@sw{\@fnsymbol}{\@fnsymbol}{\csname c@\@mpfn\endcsname}%
}%
\makeatother

\begin{document}

\preprint{AIP/123-QED}

\title[]{Estimating viscoelastic, soft material properties using a modified Rayleigh cavitation bubble collapse time}

\author{Jin Yang}
\email{jin.yang@austin.utexas.edu; Corresponding author}
\affiliation{Department of Aerospace Engineering \& Engineering Mechanics, The University of Texas at Austin, Austin, TX, USA, 78712}

 
\author{Alexander McGhee}%
\affiliation{Department of Mechanical Engineering, University of Wisconsin-Madison, Madison, WI, USA, 53706}%

\author{Griffin Radtke}%
\affiliation{Department of Mechanical Engineering, University of Wisconsin-Madison, Madison, WI, USA, 53706}%

\author{Mauro Rodriguez Jr.}%
\affiliation{ School of Engineering, Brown University, Providence, RI, USA, 02912}%

\author{Christian Franck} 
\affiliation{Department of Mechanical Engineering, University of Wisconsin-Madison, Madison, WI, USA, 53706}


\date{\today}

\begin{abstract}
Accurate determination of high strain rate (> 10$^3$ 1/s) constitutive properties of soft materials remains a formidable challenge. Albeit recent advancements among experimental techniques, in particular inertial microcavitation rheometry (IMR), the intrinsic requirement to visualize the bubble cavitation dynamics has limited its application to nominally transparent materials. Here, in an effort to address this challenge and to expand the experimental capability of IMR to optically opaque materials, we investigated whether one could use the acoustic signature of the time interval between bubble nucleation and collapse, characterized as the bubble collapse time, to infer the viscoelastic material properties without being able to image the bubble directly in the tissue. By introducing a modified Rayleigh collapse time for soft materials, which is strongly dependent on the stiffness of the material at hand, we show that, in principle, one can obtain an order of magnitude or better estimate of the viscoelastic material properties of the soft material under investigation. Using a newly developed energy-based theoretical framework, we show that for materials stiffer than 10 kPa the bubble collapse time during a single bubble cavitation event can provide quantitative and meaningful information about the constitutive properties of the material at hand. For very soft materials (i.e., shear modulus less than 10 kPa) our theory shows that unless the collapse time measurement has very high precision and low uncertainties, the material property estimates based on the bubble collapse time only will not be accurate and require visual resolution of the full cavitation kinematics. 

{  } 

\end{abstract}
 
\maketitle

\section{Introduction}
\label{sec:intro}
Recent advances in clinical and biomedical procedures involving laser and ultrasound-based modalities rely on accurate representations of the underlying material tissue properties to mitigate collateral and unwanted damage \cite{brennen2015cavitation,barney2020cavitation,tiwari2020seeded,vlaisavljevich2013image,bader2019whom,mancia2020single,mancia2021acr}. Similarly, emerging designs of personal protective equipment are beginning to model the human-material interface \cite{carlsen2021quantitative,terpsma2023head}, once again requiring accurate material models of the underlying tissue and material properties. In many of these cases, the type of material properties sought is the constitutive behavior at high strain-rates (> 10$^3$ 1/s), which until recently was challenging to acquire experimentally due to the complex and compliant nature of these materials \cite{estrada2018jmps}. Perhaps, one of the most promising, and minimally-invasive experimental techniques that recently emerged to address this need is inertial microcavitation rheometry (IMR). IMR is an inertial cavitation-based technique, specifically designed for the determination of the nonlinear viscoelastic constitutive properties of soft materials at high to ultra-high strain rates (> 10$^3$ 1/s) \cite{estrada2018jmps,yang2019semannual,yang2020eml,yang2020semannual,yang2021eml,buyukozturk2022particle,mcghee2022high}. 

While this approach works well in transparent hydrogels and translucent tissues \cite{estrada2018jmps,yang2020eml,yang2021eml,buyukozturk2022particle,mcghee2022high}, this approach has proven challenging in denser and more opaque materials such as most tissues due to its inherent requirement to optically resolve at least the first growth and collapse of the cavitation bubble. To circumvent that need, we asked whether the unique acoustic signature, derived from radially emanating shock waves during laser initiation and collapse of the cavitating bubble, could be leveraged to inform about the material properties at hand. Specifically, the detection of the characteristic collapse time of the bubble, also known as the {\it Rayleigh collapse time}, $t_c^{\text{Rayleigh}}$, can be achieved by deploying far-field transducers or hydrophones, thus removing the need to see inside the material. The classical Rayleigh collapse time is defined as the time it takes for the bubble to collapse in a given liquid, which can be derived from the continuity equation and the integration of the energy equation \cite{rayleigh1917viii}:
\begin{equation}
   t_c^{\text{Rayleigh}} = 0.91468 \ R_{\max} \sqrt{\rho/p_a}, \label{eq:t_Rayleigh}
\end{equation} 
where $\rho$ denotes the mass density of the surrounding liquid, and $p_a$ denotes the ambient pressure. The proportionality constant 0.91468 is known as the Rayleigh factor.

While the Rayleigh model and Rayleigh collapse time have historically been used to express the collapse time of a bubble within liquids \cite{lauterborn1972high}, recent experimental investigations of inertial cavitation in soft solids (i.e., hydrogels, biological tissues) have pointed toward the possibility of a similar characteristic {\it Rayleigh collapse time} within solids.   

To investigate this, we carried out a theoretical study utilizing recently published laser-induced inertial cavitation (LIC) experimental data \cite{yang2020eml,yang2021eml,mcghee2022high}. We found that a bubble's collapse time depended not only on the induced bubble size, mass density of the surrounding medium, and ambient pressure, similar to cavitation in a fluid, but also on the intrinsic viscoelastic properties of the surrounding solid. 
By deriving an energy-based framework for predicting the Rayleigh collapse time in various soft hydrogels, we were able to successfully estimate the viscoelastic material properties of these materials. The advantage of our new experimental-analytical framework presented here is that it lends itself for the estimation of the high-strain rate, viscoelastic material properties in nominally opaque materials as long as the characteristic collapse time and the maximum bubble radius can be experimentally measured or inferred. In cases where only $t_c$ can be measured, we provide an empirical scaling law for a variety of commonly used hydrogels to determine the maximum bubble radius.  
 
\begin{figure*}[t!]
\includegraphics[width = 1 \textwidth]{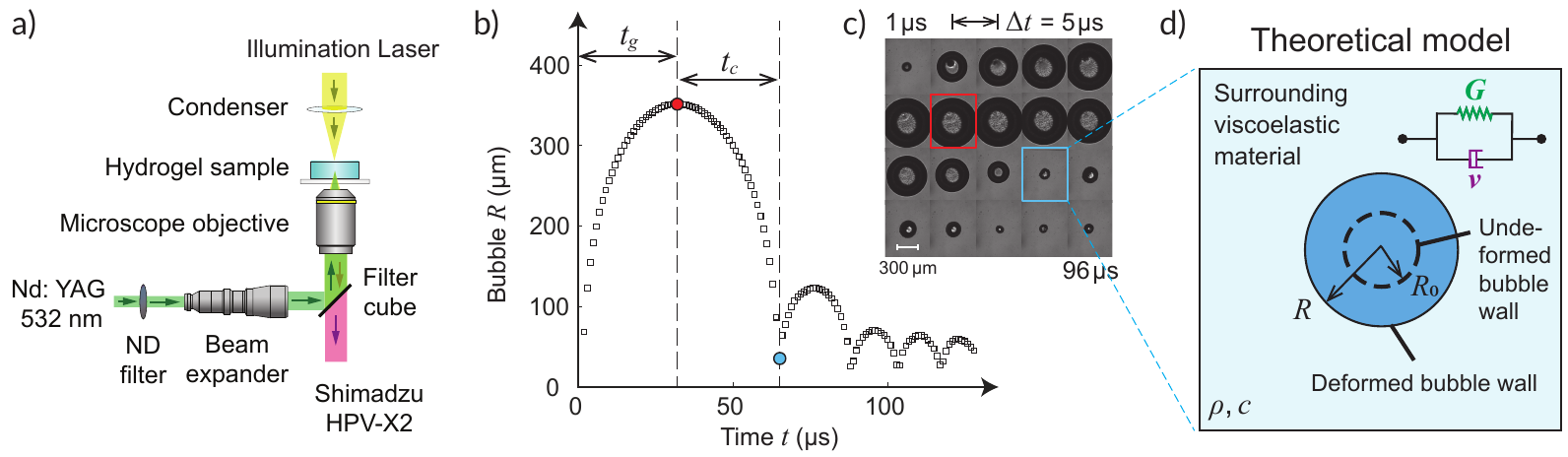} 
\caption{(a)\,The experimental set up of pulsed-laser-induced inertial cavitation (LIC) event created within a hydrogel sample. (b)\,A typical experimentally measured bubble $R$~vs.~$t$ curve for a LIC event in a soft 3\% polyacrylamide hydrogel sample. The bubble growth time $t_g$ is defined as the time between the bubble initiation and its maximum radius (red dot), and the bubble collapse time $t_c$ is defined as the time between the maximum radius and the first collapse time point (blue dot). (c)\,Twenty selected frames in the LIC event in subfigure (b) were captured by the high-speed camera. (d)\,The illustrated model is for an inertial cavitation bubble occurring within a bulk soft material modeled as a neo-Hookean Kelvin-Voigt (NHKV) viscoelastic material.} \label{fig:exp} 
\end{figure*} 
\section{Inertial cavitation in a viscoelastic soft material}
\subsection{Theoretical Analysis of Energy and Bubble Dynamics}
\label{sec:energy_theory_bulk}
%
As shown in Fig.\,\ref{fig:exp}c, here we consider an ideal model of an inertial bubble oscillation in an infinite surrounding material such that the velocity in the far-field is zero.
For an ideal, non-condensible gas inside the bubble, its internal potential energy is 
\begin{equation}
 E_{\text{PE}} = \int_{V_b} ( p_\infty - p_b ) \, \text{d}V \label{eq:BIE}
\end{equation}
%
where $V_b = 4  \pi R^3 / 3$ is the bubble volume; $p_b$ is bubble internal pressure; $p_{\infty}$ is the far-field hydrostatic pressure.

The surrounding material is assumed to be incompressible and spherically symmetric whose deformation mapping follows
\begin{equation}
    r(r_0,t) = (r_0^3 + R^3(t) - R_0^3)^{1/3} \label{eq:r}
\end{equation}
where $r$, $r_0$ are the current and reference radial positions. $R(t)$ is the time-dependent bubble radius, and $R_0$ is the radius of bubble at equilibrium. The velocity field outside the bubble can be calculated as the time derivative of (\ref{eq:r})
\begin{equation}
    v = \frac{\dot{R}R^2}{r^2}, \label{eq:v}
\end{equation}
which is also spherically symmetric. 

The kinetic energy of the surrounding material is
\begin{equation}
    E_{\text{KE}} = \int_{V_s} \frac{1}{2} \rho v^2 \text{d}V = 2 \pi \rho \dot{R}^2 R^3, 
\end{equation}
where $V_s$ is the volume of the surrounding material.

We assume that the temperature change in the surrounding material is small enough such that the change in internal energy can be neglected. 
%
%
%

The stored surface energy at the bubble wall is 
\begin{equation}
    E_{\text{SFE}} = S_b \sigma = 4 \pi R^2 \sigma,
\end{equation}
where $S_b$ is the bubble surface area and equals $4 \pi R^2$ for a spherical cavity. $\sigma$ is the surface tension coefficient at the bubble wall.

By modeling the elastic contribution of the surrounding material as a neo-Hookean (NH) hyperelastic material, its stored elastic energy can be calculated as
\begin{equation}
    E_{\text{Elastic}} = \int_{V_s} W_{\text{NH}} \, \text{d}V. \label{eq:Elastic energy}
\end{equation}%
$W_{\text{NH}}$ is the neo-Hookean strain energy density,
\begin{equation}
    W_{\text{NH}} = \frac{G}{2} \left[ \Big( \frac{r_0}{r} \Big)^4 + 2 \Big( \frac{r}{r_0} \Big)^2 - 3 \right],
\end{equation}
where $G$ is shear modulus of the surrounding material. By introducing a stretch ratio parameter $\lambda := r/r_0$, (\ref{eq:Elastic energy}) can be further simplified as (detailed derivation steps are summarized in Appendix \ref{app:E_elastic})
\begin{equation}
    E_{\text{Elastic}} = 2 \pi G \big(R^3 - R_0^3\big) \Big( \frac{2}{3} - \frac{R_0}{R} + \frac{R^2}{R^2 + R R_0 + R_0^2} \Big).
    \label{eq:E_elastic}
\end{equation}

We also assume that the surrounding material has a Newtonian viscosity. The spatial velocity gradient tensor is defined as 
\begin{equation}
    \mathbf{L} = \nabla \mathbf{v} = 
    \begin{bmatrix}
        -\frac{2 R^2 \dot{R}}{r^3} & 0 & 0 \\[4pt]
        0 & \frac{R^2 \dot{R}}{r^3} & 0 \\[4pt]
        0 & 0 & \frac{R^2 \dot{R}}{r^3}
    \end{bmatrix}.
\end{equation}
The rate of deformation tensor $\mathbf{D}$ is the symmetric part of $\mathbf{L}$,
\begin{equation}
    \mathbf{D} = \frac{1}{2} (\mathbf{L} + \mathbf{L}^{\top}).
\end{equation}

The viscous dissipation can be calculated in the following form 
\begin{equation}
    D_{\text{Visc}} = \int_{V_s} \mu \mathbf{D} : \mathbf{D} \, \text{d}V = 8 \pi \mu R \dot{R}^2, \label{eq:D_visc}
\end{equation}
where $\mu$ is the Newtonian viscosity coefficient.

The Lagrangian for the inertial cavitation system can be further defined as
\begin{equation}
    \mathcal{L} = E_{\text{KE}} - E_{\text{PE}} -  E_{\text{SFE}} - E_{\text{Elastic}}. \label{eq:Lagrangian}
\end{equation}
Following the principle of least action, we derive the Euler-Lagrange equation as \cite{johansen2016lagrangian,de2018dynamics}
\begin{equation}
    \frac{\text{d}}{\text{d}t} \Big( \frac{\partial \mathcal{L}}{\partial \dot{R}} \Big) - \frac{\partial \mathcal{L}}{\partial R} + \frac{\partial D_{\text{Visc}}}{\partial \dot{R}} = 0. \label{eq:EL}
\end{equation}

Using (\ref{eq:BIE}-\ref{eq:D_visc}) in (\ref{eq:EL}), we recover the modified Rayleigh-Plesset equation to describe inertial cavitation bubble dynamics in an infinite, Kelvin-Voigt-type viscoelastic soft material where the surrounding material is modeled as a hyperelastic neo-Hookean branch in parallel with a linear (Newtonian) viscous dash-pot (see Fig.\,\ref{fig:exp}d)  \cite{estrada2018jmps}:
\begin{equation}
    \ddot{R} R +\frac{3}{2} \dot{R}^2 = \frac{1}{\rho} \Big( p_b - \frac{2 \sigma}{R} - p_{\infty} + S_{\text{int}} \Big), \label{eq:RP}
\end{equation}
where $S_{\text{int}}$ is the stress integral of the following form
\begin{equation}
    S_{\text{int}} = - \frac{G}{2} \Big[ 5 - \Big( \frac{R_0}{R} \Big)^4 - \frac{4 R_0}{R} \Big] - \frac{4 \mu \dot{R}}{R}. \label{eq:S_integral}
\end{equation}
  
If accounting for compressibility effects in the surrounding material is desired, a modified Keller-Miksis equation (see Eq\,(\ref{eq:KM})) can be applied to account for first order compressibility effects. The modified Keller-Miksis equation (\ref{eq:KM}) will reduce to the modified Rayleigh-Plesset equation (\ref{eq:RP}) at the limit $c \to \infty$. 
\begin{equation}\label{eq:KM}
\begin{aligned}
     \left(1-\dfrac{\dot{R}}{c}\right)  R\ddot{R} & + \dfrac{3}{2}\left(1-\dfrac{\dot{R}}{3c}\right)   \dot{R}^2  \\ 
    =   \dfrac{1}{\rho} & \left(1+\dfrac{\dot{R}}{c}\right)     \left(p_{\rm b} - p_{\infty} - p_{\rm f} - \dfrac{2\gamma}{R} + S_{\text{int}} \right) \\
    + &   \dfrac{1}{\rho}\dfrac{R}{c}\dot{\overline{\left(p_{\rm b} - p_{\rm f}  - \dfrac{2\gamma}{R} + S_{\text{int}} \right)}}.
    \end{aligned}
\end{equation}

Following Estrada et al \cite{estrada2018jmps}, the bubble's internal gas composition is modeled as a mixture gas of water vapor and non-condensible, ideal gas, whose pressure is
\begin{equation}
    p_b = p_{g0} \Big( \frac{R_0}{R} \Big)^{3k} + p_{\text{v,sat}}(T_b), \label{eq:p_b}
\end{equation}
where $k$ equals the specific heat capacity ratio if we assume the bubble internal gas to be adiabatic. Then, $T_b$ is the effective averaged bubble internal temperature. $p_{\text{v,sat}}(T_b)$ is the saturation pressure of water vapor at temperature $T_b$.   $p_{g0}$ is the partial pressure of the non-condensible gas at equilibrium satisfying
\begin{equation}
    p_{g0} = p_{\infty} - p_{\text{v,sat}}(T_{\infty}) + \frac{2 \sigma}{R_0}. \label{eq:p_g0}
\end{equation}

Eqs\,(\ref{eq:RP}-\ref{eq:p_g0}) are implemented using a fifth-order explicit Dormand-Prince Runge-Kunta method with adaptive step-size control to evolve the governing equations forward in time \cite{barajas2017jasa,tzoumaka2023modeling}.

\subsection{Theoretical Analysis of the Collapse Time in an Infinite Viscoelastic Material}
\label{sec:tc_theory_bulk}
%
To develop an analytical expression for the bubble collapse time, $t_c$, we use an energy-based approach making the following assumptions \cite{kim2022energy}. We constrain ourselves to the time period up to the first bubble collapse, where we describe the dynamics of the cavitation bubble   to be inertially and elastically dominated \cite{estrada2018jmps,yang2020eml,yang2021eml}. Thus, we disregard the effects of viscosity, the bubble's internal gas contents, and temperature changes in the surrounding material, and we also assume that the total energy is conserved before the first violent collapse. Non-negligible acoustic radiation energy transfer and viscous dissipation occur primarily after the first bubble collapse, and in particular for violent collapses (i.e., Mach number > 0.1 \cite{yang2020eml,tzoumaka2023modeling}).  Given our assumptions, the total energy, $\mathcal{E}$, is then defined as 
\begin{equation}
\begin{split}
    \mathcal{E} &= E_{\text{KE}} + E_{\text{PE}} +  E_{\text{SFE}} + E_{\text{Elastic}} \\
    &= 2 \pi \rho \dot{R}^2 R^3 + \frac{4}{3} \pi R^3 p_{\infty} + 4 \pi R^2 \sigma + \\
    & \quad \ 2 \pi G \big( R^3 - R_0^3 \big) \Big( \frac{2}{3} - \frac{R_0}{R} + \frac{R^2}{R^2 + R R_0 + R_0^2 } \Big)
\end{split} \label{eq:E_total}
\end{equation}

When the bubble reaches its maximum radius, i.e., $R = R_{\text{max}}$, $\dot{R} = 0$, we also have
\begin{equation}
\begin{split}
    \mathcal{E} &= \frac{4}{3} \pi R_{\text{max}}^3 p_{\infty} + 4 \pi R_{\text{max}}^2 \sigma +  \\
    & \quad \  2 \pi G \big( R_{\text{max}}^3 - R_0^3 \big) \Big( \frac{2}{3} - \frac{R_0}{R_{\text{max}}} + \frac{R_{\text{max}}^2}{R_{\text{max}}^2 + R_{\text{max}} R_0 + R_0^2 } \Big).
\end{split} \label{eq:E_total_Rmax}
\end{equation}

Combining (\ref{eq:E_total}-\ref{eq:E_total_Rmax}), we can solve an explicit form for $\dot{R}$:
\begin{equation}
    \begin{split}
        \dot{R}^2 &= \frac{2}{3}\frac{p_{\infty}}{\rho} \Big( \frac{R_{\max}^3}{R^3} - 1 \Big) + \frac{2 \sigma}{\rho R} \Big( \frac{R_{\max}^2}{R^2} - 1 \Big)  \\
        & \quad \ + \frac{G}{\rho} \Big( \frac{R_{\max}^3}{R^3} - \frac{R_0^3}{R^3} \Big)  \Big( \frac{2}{3} - \frac{R_0}{R_{\text{max}}} + \frac{R_{\text{max}}^2}{R_{\text{max}}^2 + R_{\text{max}} R_0 + R_0^2 } \Big)  \\
        & \quad \  -  \frac{G}{\rho} \Big( 1 - \frac{R_0^3}{R^3} \Big)\Big( \frac{2}{3} - \frac{R_0}{R} + \frac{R^2}{R^2 + R R_0 + R_0^2 } \Big) 
    \end{split} \label{eq:Rdot_sq}
\end{equation}

Eq\,(\ref{eq:Rdot_sq}) can be rearranged for a differential equation for $R$:
\begin{equation}
\begin{split}
    \text{d}t &=  \Bigg[ \frac{2}{3}\frac{p_{\infty}}{\rho} \Big( \frac{R_{\max}^3}{R^3} - 1 \Big) + \frac{2 \sigma}{\rho R} \Big( \frac{R_{\max}^2}{R^2} - 1 \Big)  \\
        & \quad \ + \frac{G}{\rho} \Big( \frac{R_{\max}^3}{R^3} - \frac{R_0^3}{R^3} \Big)  \Big( \frac{2}{3} - \frac{R_0}{R_{\text{max}}} + \frac{R_{\text{max}}^2}{R_{\text{max}}^2 + R_{\text{max}} R_0 + R_0^2 } \Big)  \\
        & \quad \   -  \frac{G}{\rho} \Big( 1 - \frac{R_0^3}{R^3} \Big)\Big( \frac{2}{3} - \frac{R_0}{R} + \frac{R^2}{R^2 + R R_0 + R_0^2 } \Big)   \Bigg]^{-1/2}  \text{d}R
\end{split} \label{eq:dRdt_bulk}
\end{equation}

If $R_0/R_{\text{max}} \ll 1$, we can further integrate (\ref{eq:Rdot_sq}) from $t: 0 \rightarrow t_c$ on the left side, and  $R: R_0 \approx 0 \rightarrow R_{\text{max}}$ on the right side to obtain the approximated collapse time $t_c$ as
\begin{equation}
\begin{split}
    t_c &= \Big( \frac{2}{3} + \frac{5}{3} \frac{G}{p_{\infty}} \Big)^{- \frac{1}{2}}  \Big[  \int_{0}^{1} \frac{\xi^{3/2}}{\sqrt{1-\xi^3}} \, \text{d}\xi \Big] R_{\text{max}} \sqrt{\frac{\rho}{p_{\infty}}} \\
    &= \Big( \frac{2}{3} + \frac{5}{3} \frac{G}{p_{\infty}} \Big)^{- \frac{1}{2}} \, 0.747 \, R_{\text{max}} \sqrt{\frac{\rho}{p_{\infty}}} 
    \end{split} \label{eq:tc_bulk}
\end{equation}

One particular limit case is when the surrounding material is water, i.e., $G \rightarrow 0$, (\ref{eq:tc_bulk}) will recover the classic Rayleigh collapse time as shown in (\ref{eq:t_Rayleigh}) \cite{rayleigh1917viii}.
%

Finally, to facilitate comparison and analysis for each type of  material, we introduce the non-dimensionalized collapse time, $\psi$, as 
\begin{equation}
    \psi = t_c R_{\max}^{-1} (\rho / p_{\infty})^{-1/2}. \label{eq:psi}
\end{equation}

\section{Results}
\label{sec:results}


\begin{table*}[t!]
\caption{Summary of the utilized viscoelastic material properties and their maximum and equilibrium bubble radius, maximum circumferential stretch ratios ($\lambda_{\max}$) in LIC experiments in Section\,\ref{sec:results}} \label{tab:mat_prop}
\begin{ruledtabular}
\begin{tabular}{lccccc}
Material & $G$(kPa) & $\mu$(Pa$\cdot$s) & $R_{\max}$ &   $\lambda_{\max}$ \\ \hline
PAAM 3\% & 8.31 $\pm$ 0.43 & 0.093 $\pm$ 0.073 & 301.72 $\pm$ 26.87 &    10.3 $\pm$ 0.91 \\
PAAM 8\% & 15.09 $\pm$ 4.35 & 0.209 $\pm$ 0.180 & 330.26 $\pm$ 30.36 &   7.1 $\pm$ 0.65 \\
Agarose 0.5\% & 27.82 $\pm$ 8.56 & 0.17 $\pm$ 0.09 & 260.36 $\pm$ 24.88    &7.54 $\pm$ 0.30 \\
Agarose 1.0\% & 58.12 $\pm$ 9.38 & 0.30 $\pm$ 0.09 & 218.57 $\pm$ 10.44 &   6.57 $\pm$ 0.17 \\
Agarose 2.5\% & 114.64 $\pm$ 4.85 & 0.11 $\pm$ 0.05 & 185.55 $\pm$ 19.45   &4.32 $\pm$ 0.08 \\
Agarose 5.0\% & 282.37 $\pm$ 22.83 & 0.50 $\pm$ 0.07 & 184.41 $\pm$ 11.84 &   3.01 $\pm$ 0.15 \\
Gelatin 6\% & 17.85 $\pm$ 3.25 & 0.072 $\pm$ 0.013 & 184.0 $\pm$ 18.2  & 3.29 $\pm$ 0.36 \\
Gelatin 10\% & 36.07 $\pm$ 4.33 & 0.080 $\pm$ 0.026 & 175.9 $\pm$ 21.1   & 2.99 $\pm$ 0.36 \\
Gelatin 14\% & 108.02 $\pm$ 12.69 & 0.130 $\pm$ 0.035 & 171.0 $\pm$ 28.2   & 2.60 $\pm$ 0.19 \\
\end{tabular}
\end{ruledtabular}
\end{table*}

\begin{table*}[t!]
\caption{Estimated viscoelastic material properties using only  collapse time $t_c$ and the maximum bubble radius ($R_{\max}$) in LIC experiments} \label{tab:mat_prop_fitting_results}
\begin{ruledtabular}
\begin{tabular}{lcccccc}
\multirow{2}{*}{Material} & \multirow{2}{*}{$\psi = t_c R_{\max}^{-1} (\rho/p_{\infty})^{-1/2} $} &   \multicolumn{2}{c}{Fitted material properties using $\lbrace t_c, R_{\max} \rbrace $}   &   \multicolumn{2}{c}{Fitted material properties using $ \lbrace t_c \rbrace $ only}  \\ \cline{3-4} \cline{5-6}
  &  &   $G $(kPa)  &   $\mu$(Pa$\cdot$s) &   $G $(kPa) &    $\mu$(Pa$\cdot$s) \\ \hline
PAAM 3\% & 0.925 $\pm$ 0.013 & 1.62 $\pm$ 1.17  & 0.0233 $\pm$ 0.0156 &  1.46 $\pm$ 0.21 & 0.0230 $\pm$ 0.0027 \\
PAAM 8\% & 0.904 $\pm$ 0.017  & 3.98  $\pm$ 2.30  & 0.0468 $\pm$ 0.0205 &  3.92 $\pm$ 1.66 & 0.0479 $\pm$ 0.0124 \\
Agarose 0.5\% & 0.795 $\pm$ 0.063 & 22.72  $\pm$ 14.09  & 0.131 $\pm$ 0.0395 & 21.70 $\pm$ 6.62 & 0.1336 $\pm$ 0.0187 \\
Agarose 1.0\% & 0.705 $\pm$ 0.051 & 45.97  $\pm$ 14.08 & 0.189 $\pm$ 0.0289 & 44.26 $\pm$ 7.51 & 0.1880 $\pm$ 0.0145 \\
Agarose 2.5\% & 0.557 $\pm$ 0.017 & 120.88  $\pm$ 15.30  & 0.282 $\pm$ 0.0133 & 120.54 $\pm$ 11.93 & 0.2818 $\pm$ 0.0100 \\
Agarose 5.0\% & 0.427  $\pm$ 0.017  & 270.04 $\pm$ 28.75  & 0.372 $\pm$ 0.0129 & 268.59 $\pm$ 6.060 & 0.3714 $\pm$ 0.0027 \\
Gelatin 6\% & 1.018  $\pm$ 0.052  & 0.101 $\pm$ 0.0014 & 0.0026 $\pm$ 2.34 $\times$ $10^{-5}$ & 0.100 $\pm$ 0  & 0.0026 $\pm$ 0 \\
Gelatin 10\% & 0.729 $\pm$ 0.096  &  58.82  $\pm$ 37.25 & 0.0992 $\pm$ 0.0202 & 52.105 $\pm$ 12.785  & 0.0989 $\pm$ 0.0098 \\ 
Gelatin 14\% & 0.830 $\pm$ 0.064  & 22.46 $\pm$ 16.48  & 0.0681 $\pm$ 0.0241  & 300.898  $\pm$ 46.570  & 0.1628 $\pm$ 0.0052 \\
\end{tabular}
\end{ruledtabular}
\end{table*}

In this section we estimate the viscoelastic properties of three commonly used hydrogels using our newly derived modified Rayleigh collapse time approach based on (ii) the knowledge of $t_c$ and $R_{\max}$, or (ii) solely the measurement of the bubble collapse time, $t_c$. The general work flow of estimating the material properties is given by Algorithm \ref{alg:1}. In all instances the overall goal is to define a loss function that minimizes the least-squares difference between the estimated and experimentally measured (or informed) non-dimensional collapse time, $\psi$. 

\begin{figure*} 
\centering
\includegraphics[width = .8 \textwidth]{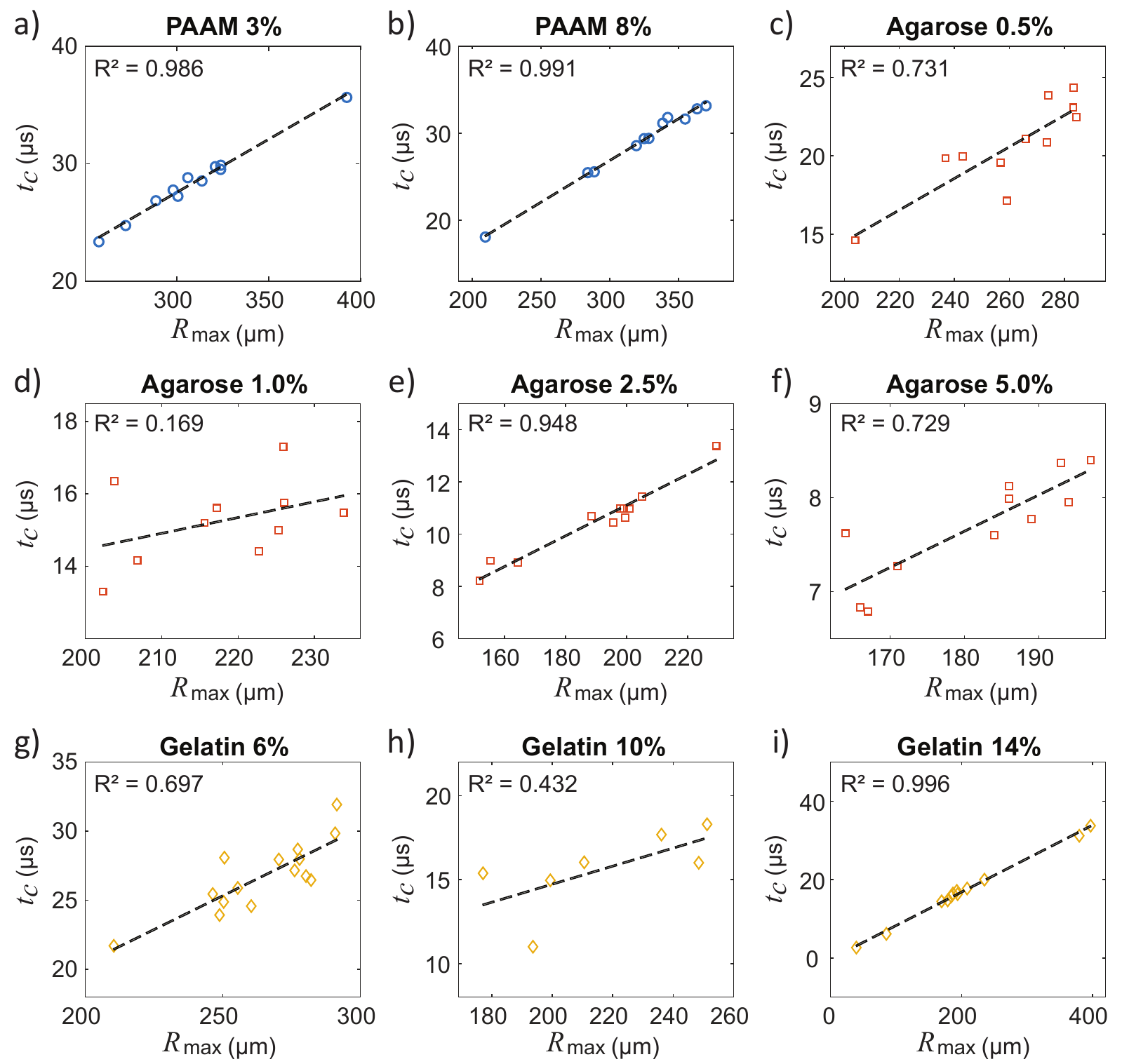}
\caption{Summary of experimentally measured laser-induced inertial cavitation (LIC) collapse time in various bulk soft materials: (a-b) 3\% and 8\% polyacrylamide (PAAM); (c-f) 0.5\%, 1.0\%, 2.5\%, and 5.0\% agarose hydrogels; (g-i) 6\%, 10\%, and 14\% gelatin hydrogels.} \label{fig:tc_bulk_each_LIC}
\end{figure*}

Since the modified Rayleigh collapse time (\ref{eq:tc_bulk}) requires knowledge of $R_{\max}$, its value must be supplied in order to calculate $\psi_{\text{exp}}$. If $R_{\max}$ is not known, it can be deduced either from the data shown in Fig.\,\ref{fig:tc_bulk_each_LIC} relating $t_c$ with $R_{\max}$ for the hydrogels presented here, found elsewhere in literature or compiled by the user themselves.  

\SetKwComment{Comment}{/* }{ */}
\RestyleAlgo{ruled}
\begin{algorithm}
\caption{Extracting unknown material viscoelastic properties using bubble collapse time}\label{alg:1}
\KwData{$t_c$, $R_{\max}$ (if known),  initially guessed shear modulus $G_i$ }
\KwResult{Viscoelastic material properties: $G$, $\mu$}
\If{$R_{\max}$  is unknown  }{
    Estimate $R_{\max}$ from Fig.\,\ref{fig:tc_bulk_each_LIC}\; 
}
Calculate $\psi_{\text{exp}} = t_c R_{\max}^{-1} \left( \rho / p_{\infty} \right)^{-1/2}$\;
Initialize $Loss > \varepsilon$, $G = G_i$\;
\While{$Loss$ $ > \varepsilon$}{
Calculate $\psi$ using (\ref{eq:tc_bulk_corrected})\;
Update $Loss = \left| \psi - \psi_{\text{exp}} \right|^2$ \;
Update $G$ using pattern search optimization scheme\;
}
Calculate estimated viscosity $\mu$ using (\ref{eq:mu_G_fit1}) or (\ref{eq:mu_G_fit2})\;
\end{algorithm}


%
In our case and for the materials presented here (Fig.\,\ref{fig:tc_bulk_each_LIC}), we quantitatively measured the bubble collapse time for each laser-induced inertial cavitation (LIC) event by analyzing the fitted   $R$~vs.~$t$ curve. The experimental measurements obtained from LIC events in polyacrylamide (PAAM), agarose, and gelatin hydrogels are shown in Fig.\,\ref{fig:tc_bulk_each_LIC}. 

\begin{figure}[h!]
\centering
\includegraphics[width = .48 \textwidth]{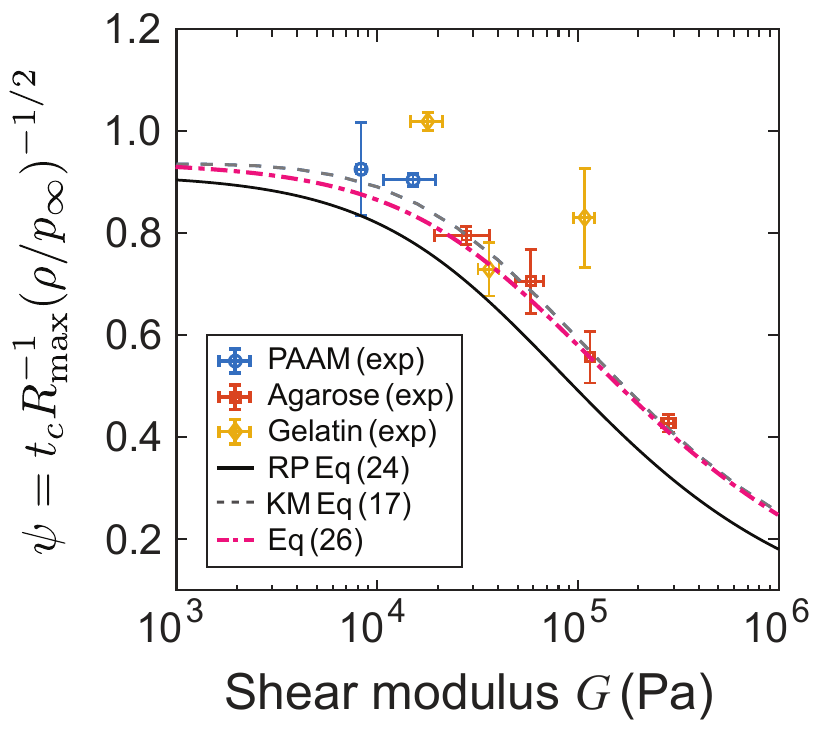}
\caption{Comparison of numerically simulated and experimentally measured bubble collapse times in bulk gel systems.} \label{fig:tc_bulk} 
\end{figure}

To facilitate comparison and analysis across all types of  material, we non-dimensionalize all experimentally measured collapse time data points from Fig.\,\ref{fig:tc_bulk_each_LIC} by using the non-dimensional parameter $\psi$ (\ref{eq:psi}).  The non-dimensionalized data points are plotted in Fig.\,\ref{fig:tc_bulk}, where the  black line represents (\ref{eq:tc_bulk}) and serves as a reference for comparison. Additionally, we performed direct numerical simulations for (\ref{eq:KM}) with different values of $R_{\max}$, which are depicted as dashed lines with corresponding values of $R_{\max}$. 

The overall shape and trend of our predictions agree well with the experimental observations, demonstrating that the collapse time of cavitation bubbles is influenced not only by their size ($R_{\max}$), the mass density and ambient pressure of the surrounding material, but also by the mechanical properties of the material itself. Specifically, we predict, consistent with experimental measurements, that a stiffer surrounding material with a higher shear modulus ($G$) results in a shorter collapse time as long as the shear modulus is sufficiently large ($>$ 10 kPa). As the shear modulus drops below 10 kPa the collapse time approaches the classical Rayleigh collapse time asymptotically and becomes nearly independent of the material's shear modulus.   
When the ratio $G/p_{\infty} \ll 1$, as indicated by (\ref{eq:tc_bulk}) and Fig.\,\ref{fig:tc_bulk}, the collapse time in a gel material follows the classic Rayleigh collapse time expressed in (\ref{eq:t_Rayleigh}).

However, as can be clearly seen from  Fig.\,\ref{fig:tc_bulk}, our expression for the modified Rayleigh collapse time in (\ref{eq:tc_bulk}) consistently underestimates the experimental measurements. This discrepancy arises due to additional factors such as material viscosity, internal gas dynamics within the bubble, and the radiation of acoustic waves, all which slow down the bubble dynamics, and lead to a prolonged first collapse time. In comparison, direct numerical simulations accurately predict the first collapse time for various hydrogel types, except for Gelatin 10\%  and 14\% hydrogels. 
To address this deviation from our prediction, we introduce a new modified Rayleigh collapse time, $t_c^{\text{corrected}}$, with an empirically derived correction factor $\theta$ accounting for some of the above mentioned effects. Its non-dimensional version, $\psi^{\text{corrected}}$ can be seen in Fig.\,\ref{fig:tc_bulk} dot-dash curve, and forms the basis for estimating the material properties shown in Table \ref{tab:mat_prop_fitting_results} using Algorithm 1.  
\begin{equation} 
    t_c^{\text{corrected}} = \theta(R_{\max}, \mu) \Big( \frac{2}{3} + \frac{5}{3} \frac{G}{p_{\infty}} \Big)^{- \frac{1}{2}} \, 0.747 \, R_{\text{max}} \sqrt{\frac{\rho}{p_{\infty}}}, \label{eq:tc_bulk_corrected}
\end{equation}
%
%



where parameters $\theta_1$ and $\theta_2$ are for  statically and dynamically crosslinked hydrogels, respectively,
\begin{align} 
    \theta_1 &= a_1 + b_1 \left[ \tanh{ c_1 (\log_{10} \mu  - d_1) } \right] \label{eq:theta1} \\
    \theta_2 &= a_2 + b_2 \left[ \tanh{ c_2 (\log_{10} \mu  - d_2) } \right] \label{eq:theta2}
\end{align}
where $\lbrace a_1, b_1, c_1 \rbrace = \lbrace 1.362, 0.335, 2 \rbrace$ and $\lbrace a_2, b_2, c_2 \rbrace = \lbrace 1.589, 0.562, 1.75 \rbrace $ are two sets of constants for  statically and dynamically crosslinked gels, respectively. Variables $d_1$ and $d_2$ are functions of maximum bubble radius which are fitted using quadratic equations as
\begin{align}
    d_1 &= -1.13 (\log_{10}R_{\max})^2 - 7.572  \log_{10}R_{\max}  - 12.83 \label{eq:d1} \\
    d_2 &= -0.2641 (\log_{10}R_{\max})^2 - 1.74  \log_{10}R_{\max}  - 3.356. \label{eq:d2}
\end{align}

Finally, it should be noted that our predictions indicate that smaller bubbles exhibit higher values of $\psi$, and there is a non-monotonic trend in $\psi$ versus $G$ when the bubble size is smaller than 80 $\upmu$m  (see SI Section S3). These deviations might be due to the increasing role of surface tension for smaller bubbles for cases when $R_{\max}$ is no longer significantly greater than $R_0$, which is an implicit assumption in the derivation of  Eq\,(\ref{eq:tc_bulk}). As this warrants a deeper investigation beyond the purpose of our initial study here, we limit the applicability of our modified Rayleigh approach to larger bubbles with nominal radii above 80 $\upmu$m.

\subsection*{Estimation of Viscoelastic Material Properties using the modified Rayleigh collapse time}
\begin{figure}[h!]
\centering
\includegraphics[width = .45  \textwidth]{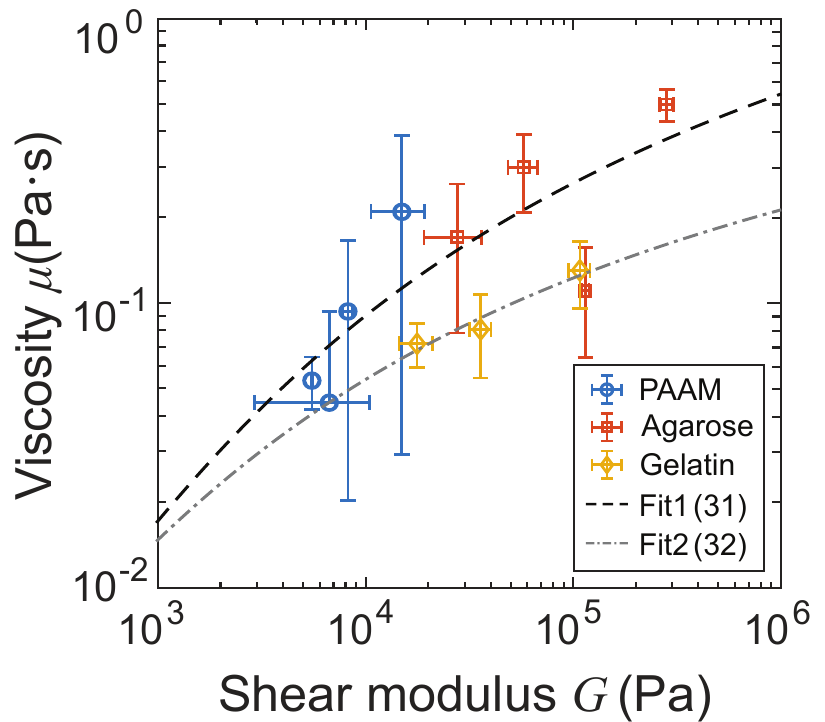}
\caption{Fitted material viscosity ($\mu$) vs. shear modulus $G$ from laser-induced inertial cavitation (LIC) experimental measurements \cite{yang2020eml,yang2021eml,mcghee2022high}. Fit1 and Fit2 are for statically (i.e., PAAM and agarose) and dynamically (i.e., gelatin) crosslinked hydrogels, respectively.} \label{fig:mat_prop}
\end{figure}

As described in Algorithm 1, the material shear modulus $G$ is estimated via a least-squares minimization procedure of the non-dimensional collapse time, $\psi$ (Steps 1-11), after which the material viscosity is calculated using the following two equations obtained by fitting the data presented in Fig.\,\ref{fig:mat_prop} for statically (i.e., PAAM and agarose) and dynamically (i.e., gelatin) crosslinked hydrogels, respectively (Step 12), 
\begin{align}
    \mu^{\text{Fit1}} &= 20.70 \times 10^{ -9.48/\text{log}_{10} (G+156.96)  } \label{eq:mu_G_fit1} \\
    \mu^{\text{Fit2}} &= 3.42 \times 10^{ -7.24/\text{log}_{10} (G+111.90)  } \label{eq:mu_G_fit2}
\end{align} 

Using the experimentally measured collapse time and $R_{\max}$ values shown in Table \ref{tab:mat_prop}, all estimated material properties using our new modified Rayleigh collapse time approach (i.e., Algorithm 1) are summarized in Table \ref{tab:mat_prop_fitting_results} for the case (i) when both $t_c$ and $R_{\max}$ are known, and (ii) if only $t_c$ is available but $R_{\max}$ can be deduced from a prior database or correlation as presented in Fig.\,\ref{fig:tc_bulk_each_LIC}.

\section{Discussion and Conclusion}
\label{sec:summary}
Laser-induced cavitation (LIC) affords the user the flexibility of generating micron to centimeter-sized bubbles in a spatially controllable and repeatable fashion. Estimating the material properties using LIC such as with IMR requires the optical resolution of the resulting bubble kinematics, which limits its current application to nominally transparent materials. The emanation of two clearly distinguishable acoustic shocks, one at bubble initiation and one at first bubble collapse, allows recording of a well-defined bubble collapse time, which we have shown is strongly dependent on the stiffness of the surrounding material and can be cast in the form of a modified Rayleigh collapse time (\ref{eq:tc_bulk}). The large amplitude of both of these pressure waves allows them to be easily recorded in the far field, which makes this approach, as described by Algorithm 1, attractive for estimating high strain rate material properties of optically opaque materials.

We find that the estimated viscoelastic properties of three commonly used hydrogels are similar whether using the knowledge of $t_c$ and $R_{\max}$, or solely using the measurement of the bubble collapse time $t_c$. When comparing the results obtained via Algorithm 1 to the material properties obtained from IMR as a ground truth estimate, we find that the results are close except for 6\% and 14\%  gelatin hydrogel cases. This is to be expected since the 6\% and 14\% gelatin concentration results also deviate from our theoretical predictions in Fig.\,\ref{fig:tc_bulk}. We also find that the relative errors of the estimated material properties are less for stiffer hydrogels. Generally speaking, when the shear modulus is $G >$ 10 kPa, the relative errors of the shear modulus $G$ and viscosity $\mu$ are less than 20\% and 40\%, respectively.

Finally, we  conclude with a few limitations and observations. First, estimating a material's mechanical properties based on its characteristic Rayleigh collapse time is most accurate for larger (i.e., $R_{\max} > 80$ $\upmu$m) bubbles in stiffer materials (shear modulus $>$ 10 kPa). For smaller bubbles, the scaling behavior is more complex (Fig.\,\ref{fig:tc_bulk}) as predicted by (\ref{eq:tc_bulk}) and will require additional treatment beyond the scope of this work. Furthermore, if the material is softer than 10 kPa, the modified Rayleigh collapse time approaches the collapse time in water and becomes nearly independent of the shear modulus of the materials, significantly limiting our model's ability to provide meaningful material information. Second, our current energy-based framework does not explicitly account for the effect of material viscosity and dissipation but rather offers an empirical solution via a correction factor and a phenomenological fit for its determination. Future studies can improve upon this by potentially accounting for this within the framework itself. Finally, the choice of a hyperelastic Neo-Hookean solid for the elastic contribution of the material model can be modified by the user to perhaps more appropriately represent the material. As can be seen from Fig.\,\ref{fig:tc_bulk}, that this particular choice in constitutive model agrees well with the three hydrogels examined here.  

In sum, we derived and presented a new modified Rayleigh collapse time, $t_c$, for soft solids analogous to what has been derived for pure fluids. As $t_c$ strongly depends on the shear modulus of the surrounding material, we showed that knowledge of $t_c$ along with an estimate of $R_{\max}$, the maximum bubble radius, during laser-induced inertial cavitation, can provide a meaningful estimate of the underlying high strain-rate material properties. 

\section*{Acknowledgements}
We gratefully acknowledge support from the US Office of Naval Research under PANTHER award number N000142112044 through Dr. Timothy Bentley.

\appendix
\section{Appendix A: Derivation of Eq\,(\ref{eq:E_elastic}) }
\label{app:E_elastic}
%
After introducing stretch ratio parameter $\lambda := r/r_0$ and using Eq(\ref{eq:r}), we can re-write $r$ and d$r$ in terms of $\lambda$:
\begin{align}
        r &= \big( R^3 - R_0^3 \big)^{1/3} \lambda \big( \lambda^3 - 1 \big)^{-1/3} \\
        \text{d}r &= - \big( R^3 - R_0^3 \big)^{1/3} \big( \lambda^3 - 1 \big)^{-4/3} \text{d}\lambda.
\end{align}

Eq(\ref{eq:Elastic energy}) can be further simplified as
\begin{equation}
    \begin{split}
      E_{\text{Elastic}} &= \int_{R}^{\infty} 4 \pi r^2 \frac{G}{2} \left[ \Big( \frac{r_0}{r} \Big)^4 + 2 \Big( \frac{r}{r_0} \Big)^2 - 3 \right] \, \text{d}r \\
      &= 4 \pi (R^3 - R_0^3) \int_{1}^{R/R_0} \frac{\lambda^2 W_{\text{NH}}(\lambda)}{(\lambda^3-1)^2} \, \text{d}\lambda \\
      &=  2 \pi G \big(R^3 - R_0^3\big) \Big( \frac{2}{3} - \frac{R_0}{R} + \frac{R^2}{R^2 + R R_0 + R_0^2} \Big).
    \end{split}
\end{equation}

%
%

\section*{References}
\bibliographystyle{unsrt}
\bibliography{reference}

\end{document}


\title{SUPPLEMENTARY MATERIAL of \\
``Estimating viscoelastic, soft material properties using a modified Rayleigh
cavitation bubble collapse time"}%

\author{Jin Yang}
\email{jin.yang@austin.utexas.edu; Corresponding author}
\affiliation{Department of Aerospace Engineering \& Engineering Mechanics, The University of Texas at Austin, Austin, TX, USA, 78712} 

\author{Alexander McGhee}  
\affiliation{Department of Mechanical Engineering, University of Wisconsin-Madison, Madison, WI, USA, 53706}

\author{Griffin Radtke}  
\affiliation{Department of Mechanical Engineering, University of Wisconsin-Madison, Madison, WI, USA, 53706}

\author{Mauro Rodriguez Jr.} 
\affiliation{School of Engineering, Brown University, Providence, RI, USA, 02912}

\author{Christian Franck}  
\affiliation{Department of Mechanical Engineering, University of Wisconsin-Madison, Madison, WI, USA, 53706}

\date{\today} 
\maketitle
\onecolumngrid

\section{Summary of all used IMR data points}

\begin{figure}[h!]
\begin{center} 
\includegraphics[width=0.42 \textwidth]{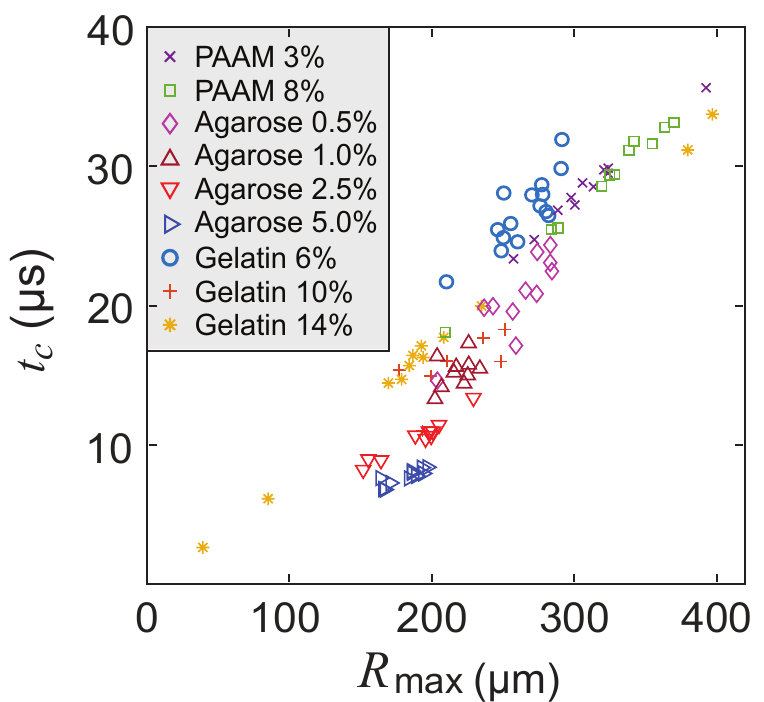} 
\end{center}
\caption{Summary of all used IMR data points from three commonly used hydrogels \cite{yang2020eml,yang2021eml,mcghee2022high}.}
\label{fig:data_summary}
\end{figure}

All used pulsed-laser-induced inertial microcavitation bubble collapse time $t_c$ and maximum bubble radii are summarized in Fig.\,\ref{fig:data_summary}.

\section{Fitted empirical relation between maximum circumferential stretch ratio and the surrounding material's shear modulus}

\begin{figure}[h!]
\begin{center} 
\includegraphics[width=0.42 \textwidth]{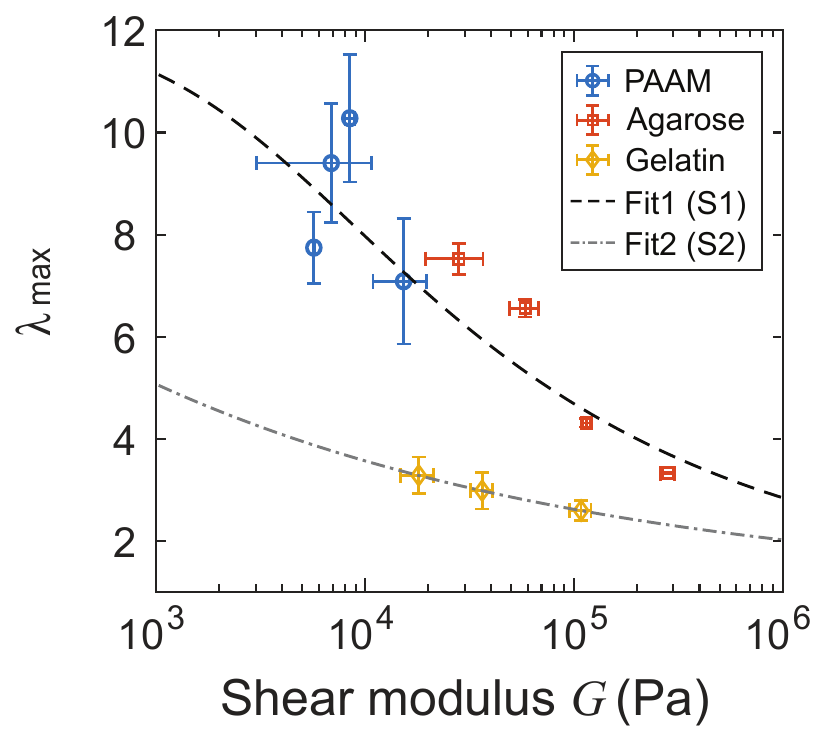} 
\end{center}
\caption{Fitted maximum circumferential stretch ratio ($\lambda_{\max} =
R_{\max}/R_0$) vs. shear modulus $G$ from laser-induced inertial cavitation (LIC) experimental measurements \cite{yang2020eml,yang2021eml,mcghee2022high}.}
\label{fig:lambda_fit}
\end{figure}

We further fit the maximum circumferential stretch ratio ($\lambda_{\max} = R_{\max}/R_0$) as functions of the surrounding material's shear modulus using two phenomenological models for hydrogels with static crosslinks (Fig1; i.e., polyacrylamide (PAAM) and agarose) and dynamic crosslinks (Fit2; i.e., gelatin), as shown in (\ref{eq:lambda_max_G_fit1},\ref{eq:lambda_max_G_fit2}). The corresponding fitting results are shown in Fig.\,\ref{fig:lambda_fit}.
\begin{align}
    \lambda_{\max}^{\text{Fit1}} &= 1 + 120.50 \times 10^{-0.302 \times \text{log}_{10} (G + 2609.1) } \label{eq:lambda_max_G_fit1} \\
    \lambda_{\max}^{\text{Fit2}} &= 1 + 16.07 \times 10^{-0.199 \times \text{log}_{10} (G + 6.130) } \label{eq:lambda_max_G_fit2}  
\end{align}
%

\section{Comparison of numerically simulated and experimentally
measured bubble collapse times in bulk gel systems}

\begin{figure}[h!]
\begin{center} 
\includegraphics[width=0.9 \textwidth]{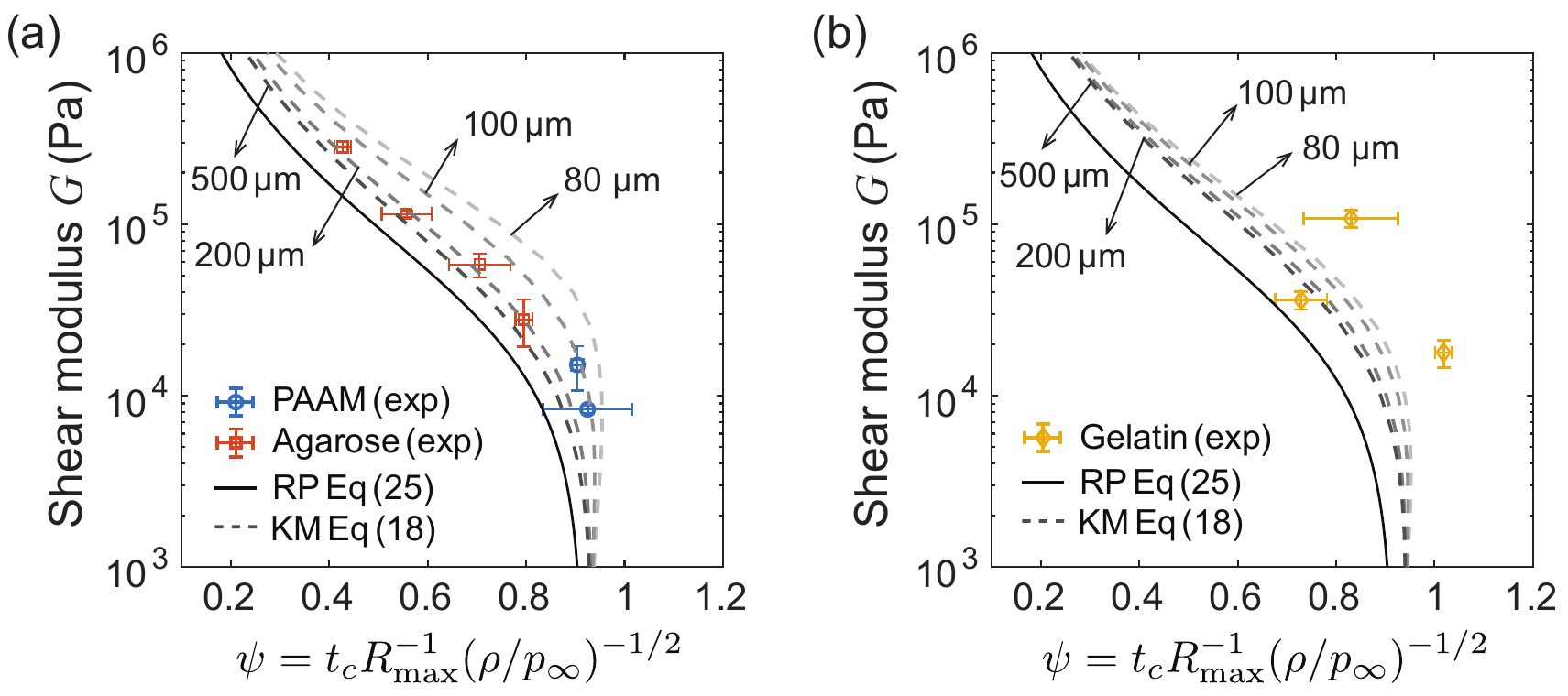} 
\end{center}
\caption{Comparison of numerically simulated and experimentally
measured bubble collapse times in infinite bulk gel systems: (a) staticly
crosslinked and (b) dynamic crosslinked hydrogels.}
\label{fig:psi_vs_G}
\end{figure}

We non-dimensionalize all experimentally measured collapse time data points in Fig. 3 in the main text by dividing them by $R_{\max} (\rho / p_{\infty})^{1/2} $. We also perform numerical simulations to solve Keller-Miksis equation (17) with different values of $R_{\max}$, which are depicted as dashed lines in Fig.\,\ref{fig:psi_vs_G} with corresponding values of $R_{\max}$. We find that there is a non-monotonic trend in $\psi$ versus G when the bubble size is smaller than 80 $\upmu$m.

\section{Coefficient $\theta$ to include all other effects due to viscosity and bubble size}

As discussed in the main text section ``\textbf{Estimation of Viscoelastic Material Properties Using the Modified Rayleigh Collapse Time}", we introduced a correction coefficient $\theta$ ($\theta_1$ for static crosslinks, i.e., polyacrylamide (PAAM) and agarose hydrogels; and $\theta_2$ for dynamic crosslinks, i.e., gelatin hydrogels) in front of non-dimensionalized collapse time $\psi$. Here we show the goodness of our fittings in (27-30) in the main text by comparing our numerical simulated results in Fig. 3.

\begin{figure}[h!]
\begin{center} 
\includegraphics[width=0.9 \textwidth]{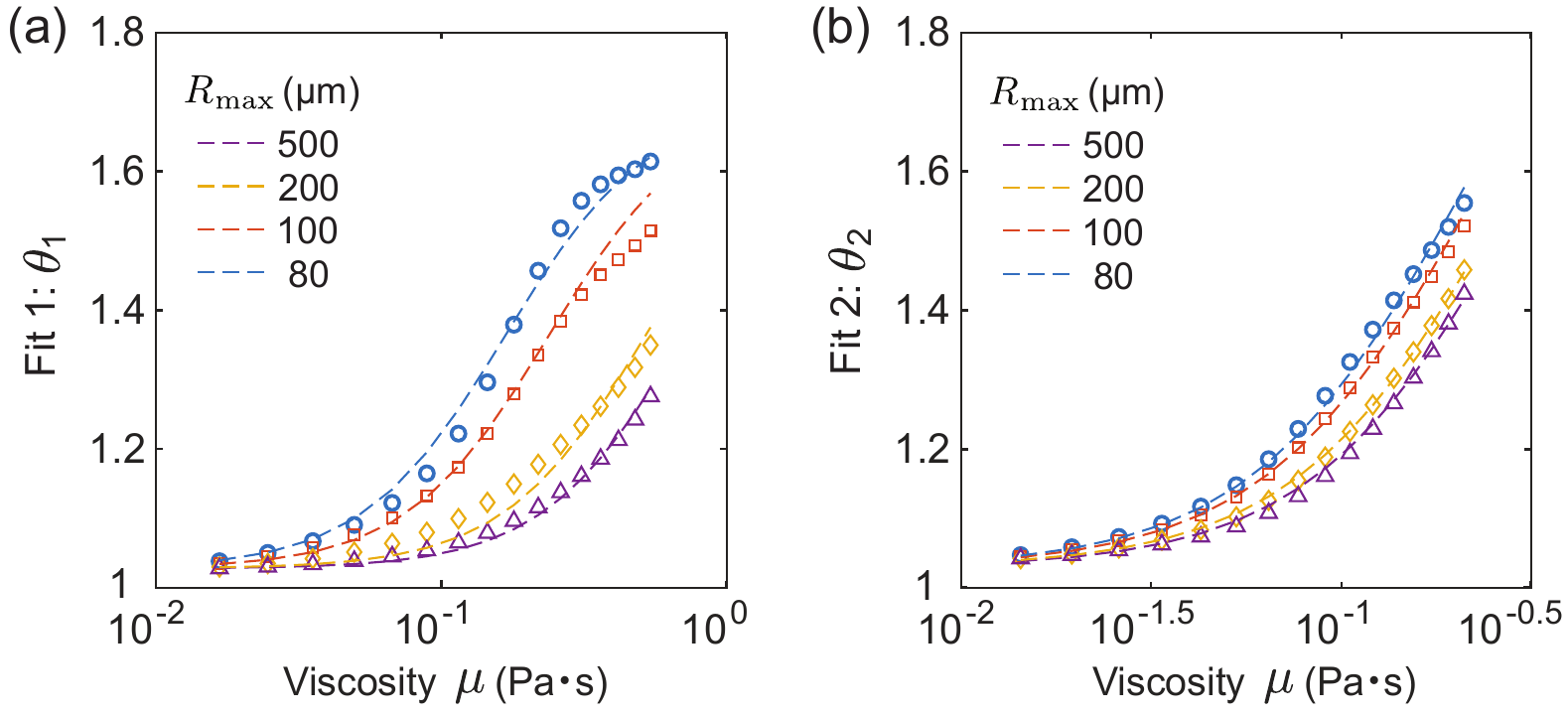} 
\end{center}
\caption{Correction coefficients $\theta_1$ and $\theta_2$ for  statically (i.e., PAAM and agarose) and dynamically (i.e., gelatin) crosslinked hydrogels are numerically calculated (hollow markers)
and then empirically fitted (dashed lines).}
\label{fig:theta}
\end{figure}

\bibliography{reference}